\documentclass[aps,prl, superscriptaddress,twocolumn]{revtex4}
\usepackage{feynmp}
\usepackage{amsmath}
\usepackage{amssymb}
\usepackage{epsfig}
\usepackage{graphicx}
\usepackage{subfigure}
\usepackage{hyperref}
\usepackage{bbm}
\usepackage{color}
\usepackage{longtable}
\usepackage{booktabs}
\usepackage{multirow}
\usepackage{array}
\newcommand{\tabincell}[2]{\begin{tabular}{@{}#1@{}}#2\end{tabular}}

\newcommand{\Rmnum}[1]{\expandafter\@slowromancap\romannumeral #1@}
\allowdisplaybreaks[3]
\begin{document}

\title{Comment on  ``Emergence of Nodal-Knot Transitions by Disorder"}

\author{Jing-Rong Wang}
\altaffiliation{Corresponding author: wangjr@hmfl.ac.cn}
\affiliation{High Magnetic Field Laboratory of Anhui Province,
Chinese Academy of Sciences, Hefei 230031, China}

\date{\today}

\maketitle

%%%%%%%%%%%%%%%%%x%%%%%%Main Body%%%%%%%%%%%%%%%%%%%%%%%%%%%%%%%%%%%%%

In a recent paper \cite{Gong24}, Gong \emph{et al.} studied the disorder effects in nodal-knot semimetal through Wilson
momentum-shell renormalization group (RG) method. They stated that various nodal-knot transitions emerge driven by disorders.
However, we notice that there are serious problems in this paper.

\textbf{\emph{1. Incorrect momentum space for RG analysis.}}

 For a fermion system, the RG analysis
should be performed in the momentum space around the Fermi surface \cite{Shankar94, Polchinski92, Senthil09}.
In the RG analysis, fermion momenta should be scaled towards the Fermi surface \cite{Shankar94, Polchinski92, Senthil09}.
Codimension $d_{c}$ is an import concept since it determines  number of momentum components which
are scaled in the RG analysis,  and it is defined as
\begin{eqnarray}
d_{c}=d-d_{\mathrm{FS}},
\end{eqnarray}
where $d$ is the dimension of the system and $d_{\mathrm{FS}}$ is the dimension of the Fermi surface \cite{Senthil09}.
For three-dimensional (3D) Dirac semimetal, the Fermi surface is 0D point.
Accordingly, codimension for 3D Dirac semimetal is three. Thus, the RG analysis
is performed in the momentum space with three momentum components around the Dirac point \cite{Goswami11, Roy16}. Concretely, the RG transformations are applied
for the three momentum components and momentum shell is imposed for these components.  However, for 3D nodal line
semimetal, the Fermi surface is 1D line. Accordingly, the codimension for nodal line semimetal is two. The RG analysis
should be performed in the momentum space with two momentum components which are perpendicular to the nodal line \cite{Huh16, Sur16, Roy17, WangYuXuan17, Han17, WangLiuZhangWan20}. Thus, the RG transformations
are only applied for these two momentum components, and the momentum shell is also imposed for these two components.

In Ref.~\cite{Gong24}, the author considered nodal knot, nodal link semimetals \emph{etc.} which have 1D line Fermi surface. The codimension of these semimetals is two.
However, they did't not perform the RG analysis in the momentum space around the Fermi surface, but in the three dimensional momentum space around the
point $k=0$. In Ref.~\cite{Gong24}, the momenta are not scaled towards the Fermi surface, but scaled towards the point $k=0$.
Thus, the power counting and calculation of loop corrections in Ref.~\cite{Gong24} both are unreliable. There is an obvious mistake for their RG equations.
According to the first term of r.h.s. of Eqs.~(7) and (8) in Ref.~\cite{Gong24}, the disorder is irrelevant to tree-level.
However, it is well known that disorder is marginal to tree-level in nodal line semimetal \cite{WangYuXuan17}.

\textbf{\emph{2. Invalid criterion for phase transition.}}

In Ref.~\cite{Gong24}, they replaced the constant parameters $m_{1}$, $m_{2}$ \emph{etc.} in the Hamiltonian with the RG flows $m_{1}(l)$, $m_{2}(l)$ \emph{etc.}  with
$l$ being the RG running parameter, then they analyzed the knot Wilson loop integral $W$ for this $l$ dependent Hamiltonian and defined it as $W(l)$.
When change of $m_{1}(l)$ or $m_{2}(l)$ with increasing of $l$ is large, $W(l)$ jumps from a constat to another constant.  They took jump of $W(l)$ as the criterion
for the phase transition from a nodal-knot configuration to another configuration.

They took the initial value $m_{1}(l=0)=0$, $m_{2}(l=0)=0$, and found that
$m_{1}(l)$ or $m_{2}(l)$ changes from $0$ gradually with increasing of $l$ under the influence of disorder.
They showed that $W(l)$ holds the original value when the changes of $m_{1}(l)$, $m_{2}(l)$ are small, but jumps to a new value when the change of
$m_{1}(l)$ or  $m_{2}(l)$ is large enough. Then they regarded these results as the nodal-knot phase transition driven by disorder.

However, it is easy to find that the criterion for phase transition employed in Ref.~\cite{Gong24} is invalid. Neglecting various disorder, the
RG equations  for $m_{1}(l)$ and $m_{2}(l)$ shown in Eq.~(6) of Ref.~\cite{Gong24} become
\begin{eqnarray}
\frac{dm_{1}}{dl}=m_{1}, \label{Eq:RGEm1}
\\
\frac{dm_{2}}{dl}=m_{2}, \label{Eq:RGEm2}
\end{eqnarray}
which are RG equations for $m_{1}(l)$ and $m_{2}(l)$ in clean system. The solutions for Eqs.~(\ref{Eq:RGEm1}) and (\ref{Eq:RGEm2}) are
\begin{eqnarray}
m_{1}(l)=m_{1}(0)e^{l},
\\
m_{2}(l)=m_{2}(0)e^{l}.
\end{eqnarray}
If we take $m_{1,2}(0)$ with a small positive value, $m_{1,2}(l)$ increases gradually from this value and approaches positive infinity when $l\rightarrow\infty$;
If we take $m_{1,2}(0)$ with a small negative value, $m_{1,2}(l)$ decreases gradually from this value and approaches negative infinity when $l\rightarrow\infty$.
Accordingly, the corresponding $W(l)$ holds on if the change of $m_{1,2}(l)$ is small, but jumps to a new value if the change of $m_{1,2}(l)$ is large enough.

As an concrete example, for $p=q=2$ where $p$ and $q$ are defined in Eq.~(3) of Ref.~\cite{Gong24}, taking $m _{1}(0)=-10^{-6}$, $m_{2}(0)=0$, we can find that $W(l)$ is still equal to 2 if  decrease of $m_{1}(l)$ is small, but
finally jumps to $0$ when $m_{1}(l)$ is small enough. According to the criterion in Ref.~\cite{Gong24}, it represents a phase transition from Hopf link semimetal
to unlink nodal line semimetal.

As another concrete example, for $p=q=3$, taking $m_{1}(0)=0$, $m_{2}(0)=10^{-6}$, $W(l)$ is still equal to 6 if increase of $m_{2}(l)$ is small, but finally jumps
to $3$ when $m_{2}(l)$ is large enough. According to the criterion in Ref.~\cite{Gong24}, it represents a phase transition from valknut nodal-knot semimetal to trefoil nodal-knot
semimetal.

\emph{\textbf{Therefore, we can find that the various phase transitions stated in Ref.~\cite{Gong24} even emerge in clean nodal-knot semimetal system!}} This reflects that taking jump of $W(l)$ as
transition of nodal-knot semimetal from one configuration to another is completely misleading.

\textbf{\emph{3. $W(l)$ is an ill-defined quantity.}}

In Ref.~\cite{Gong24}, RG running parameter $l=0$ is corresponding to $k=\Lambda$ where $\Lambda$ is momentum cutoff, $l\rightarrow\infty$ corresponds to $k\rightarrow0$,
and general $l$ is corresponding to a general local momentum $k=\Lambda e^{-l}$. Thus, $W(l)$ is actually a quantity corresponds to a local momentum scale. However, the knot Wilson loop integral
is a global property of the system. It indicates that $W(l)$ is an ill-defined meaningless quantity.

In summary, in Ref.~\cite{Gong24}, the RG analysis is not performed in the momentum space around Fermi surface but in an incorrect momentum space,
accordingly the RG equations are unreliable. Additionally, the criterion for phase transition
employed in Ref.~\cite{Gong24} is invalid. Thus, the conclusions in this paper are unfounded.

\begin{table*}[htbp]
\caption{Details in the references for supporting the scientific statements.
\label{Table:SummaryRGResult}}
\begin{center}
\begin{tabular}{|l|l|}
\hline\hline    Scientific statements &  Details in the references
\\
\hline          \tabincell{l}{For a fermion system, the RG analysis should be performed \\
in the momentum space around the Fermi surface.  \\
In the RG analysis, fermion momenta should be scaled \\
 towards the Fermi surface. }  &
\tabincell{l}{$\star$. 1st paragraph of page 6, 2nd paragraph of page 7,   \\
\hspace{0.3cm}  Table II of page 40, 1st paragraph of page 59 in Ref.~\cite{Shankar94}.  \\
$\star$. 1st paragraph of page 14 in Ref.~\cite{Polchinski92}. \\
$\star$. 1st paragraph of page 1 in Ref.~\cite{Senthil09}. }
\\
\hline \tabincell{l}{For a 3D system with 1D line Fermi surface, the RG  \\
analysis  should be performed in the momentum \\
space with
two momentum components which are  \\
perpendicular to the nodal line.\\
The momenta should be scaled towards the nodal line, \\
but not towards the point $k=0$. } & \tabincell{l}{$\star$. 3rd, 4th paragraphs of page 1 in Ref.~\cite{Senthil09}.\\
$\star$. 2nd, 3rd paragraphs of page 2 in Ref.~\cite{Huh16}.\\
$\star$. 6th paragraph of page 2, 2nd paragraph of page 4, \\
\hspace{0.4cm}2nd paragraph of page 6 in Ref.~\cite{Sur16}. \\
$\star$. 3rd paragraph of page 2 of main text, and 1st,  2nd   \\
\hspace{0.3cm} paragraphs of page 4 in Supplementary Materials in Ref.~\cite{Roy17}.\\
$\star$. 3rd paragraph of Section II in Ref.~\cite{WangYuXuan17}. \\
$\star$. 4th paragraph of page 1, 4th, 5th paragraphs of page 3 \\
\hspace{0.4cm}in Ref.~\cite{Han17}.\\
$\star$. 1st paragraph of Appendix A~2, 1st paragraph of \\
\hspace{0.4cm}Appendix B in Ref.~\cite{WangLiuZhangWan20}. }
\\
\hline \tabincell{l}{For  3D nodal line semimetal, the disorder is marginal \\
 to tree-level. }  & \tabincell{l}{$\star$.  3rd, 4th paragraphs of Section I, 1st paragraph of page 5, \\
\hspace{0.4cm}1st paragraph of Section VI in
 Ref.~\cite{WangYuXuan17}.}
\\
\hline \hline
\end{tabular}
\end{center}
\end{table*}

\end{document}